\documentclass[prb,preprint]{revtex4}
\usepackage [dvips]{graphicx}
\usepackage{amssymb}

\begin{document}

\title{
Giant magnetoimpedance in Vitrovac$^{\textrm{\scriptsize\textregistered}}$ amorphous 
ribbons over [0.3-400 MHz] frequency range}

\author{J. Gieraltowski, A. Fessant, R. Valenzuela \P  \hspace{1mm} and  C. Tannous}
\affiliation{UBO, LMB-CNRS UMR 6135, BP: 809 Brest CEDEX, 29285 France \\
\P Instituto de Investigaciones en Materiales, Universidad 
Nacional Aut\'{o}noma de Mexico, P.O. Box 70-360, Coyoacan, Mexico D.F., 04510, Mexico}

\begin{abstract}
Giant magneto impedance (GMI) effect for as-cast 
Vitrovac$^{\textrm{\scriptsize\textregistered}}$ amorphous ribbons (Vacuumschmelze, 
Germany) in two configurations (parallel and normal to the ribbon axis) is 
studied over the frequency range [0.3-400 MHz] and under static magnetic 
fields  -160 Oe $< H_{dc} < $+160 Oe. A variety of peak features and GMI 
ratio values, falling within a small field range, are observed and 
discussed.
\end{abstract}


\maketitle

The giant magneto-impedance effect (GMI) in amorphous ribbons and thin films 
has become a topic of growing interest for a wide variety of prospective 
applications in storage information technology and sensors possessing high 
sensivity and fast response \cite{mohri,panina}. 

Magneto impedance effect (MI) consists in change of impedance introduced by 
a low-amplitude alterning-current (\textit{ac}) flowing through a magnetic conductor 
under application of a static magnetic field $H_{dc}$ (usually applied in the 
plane along (LMI) or perpendicular (TMI) to the direction of the probe 
current). The origin of this behavior is related to the magnetic relative 
permeability $\mu_{r}$ value of materials having the right value and direction 
of magnetic anisotropy field $H_{k}$ that yields a given GMI ratio and 
profile versus $H_{dc}$ field and frequency $f$. For instance, in the LMI case, 
MI profile versus $H_{dc}$ field exhibits either a single- or a double-peak, 
for easy magnetization axis (anisotropy axis) parallel or perpendicular to 
the current direction, respectively. The same behaviour is also observed 
when frequency $f$ is varied.

For planar geometry (ribbon and thin films) and in-plane uniaxial magnetic 
anisotropy \cite{panina, atkinson}, a large GMI ratio (dependent on
 $\sqrt{\mu_{r}}$ is obtained with the easy magnetization 
direction transverse to the longitudinal axis of conductor possessing along 
this direction a ``transverse'' permeability $\mu_{rT}$. Thus, it seems that a 
large transverse permeability  $\mu_{rT}$ resulting from the very small magnetic 
anisotropy (which is inherent to amorphous alloys) is necessary to give a 
strong GMI effect. But it is also shown \cite{panina,atkinson} that the maximum GMI value is 
not really dependent on the magnetic anisotropy but strongly on the magnetic 
softness, and, in the case of amorphous alloys, relies upon a small but 
well-defined anisotropy in the transverse direction \cite{fessant}. 

The distribution of magnetic anisotropy can also alter the GMI effect. It 
can be decomposed into longitudinal and transverse components along the long 
and short direction of the ribbon respectively and plays an important role 
in the GMI effect \cite{krauss, pirota, sommer}. 

The purpose of this work is to examine experimentally the GMI effect in LMI 
and TMI configurations with different magnetic anisotropy distributions in 
order to discriminate among all these components. 

The GMI measurements were carried out on as-cast amorphous ribbons with nominal composition: 
Co$_{66}$Fe$_{4}$Mo$_{2}$B$_{16}$Si$_{12}$ (Vitrovac$^{\textrm{\scriptsize\textregistered}}$ 6025). 
This Cobalt rich metallic glass alloy is interesting because of 
its relative permeability that can reach as high as 100,000. 

Samples (2mm x 15mm x 30 $\mu $m) were cut parallel (CP) as well as 
transverse (CT) to the ribbon long axis from an as-cast commercial band 
having the anisotropy axis along the ribbon axis. The samples obtained show 
the easy axis oriented only along the long geometrical axis of the samples 
(confirmed by hysteresis loop measurements) and different levels of 
anisotropy dispersion. For the transverse-cut sample (CT) the anisotropy 
axis is altered by strong demagnetizing field $H_{dem}$ present in this 
configuration and exhibits an in-plane rotation with respect to its original 
orientation. 

Measurements of the GMI ratio were carried out by applying a field 
$H_{dc}$ parallel or perpendicular to the long sample axis using a novel 
broad band measurement method described elsewhere \cite{fessant}. All measurements were 
made at room temperature, in the frequency range [0.3 - 400 MHz], under 
$H_{dc}$ fields -160 Oe $ < H_{dc} <  $ +160 Oe and low-amplitude \textit{ac} 
current (0.1 mA). The MI ratio was determined by the expression:
$$
\frac{\Delta Z}{Z} = \left| {\frac{Z(H,f) - Z(H_{\max } ,f)}{Z(H_{\max } 
,f)}} \right|
$$
where$ H_{max}$ is the maximum value of $H_{dc}$ (that is 160 Oe). 

Taking into account the easy axis orientations of the CT and CP samples 
measured in LMI and TMI configurations, a single-peak behavior is expected. 
However, the presence of a transverse anisotropy component and a very low 
$H_{dem}$ field in the LMI case results in a double-peak behavior. Thus, the 
GMI ratio versus $H_{dc}$ is the typical curve \cite{fessant} doubly peaking at $H =\pm $ 
($H_{kT }+H_{dem})$ close to the transverse component of the anisotropy 
field $H_{kT}$with a separation of about 9 Oe between peaks (Fig.1) for the 
CP sample. 

\begin{figure}[h!] 
\centering
\includegraphics[angle=0, width=13.5cm]{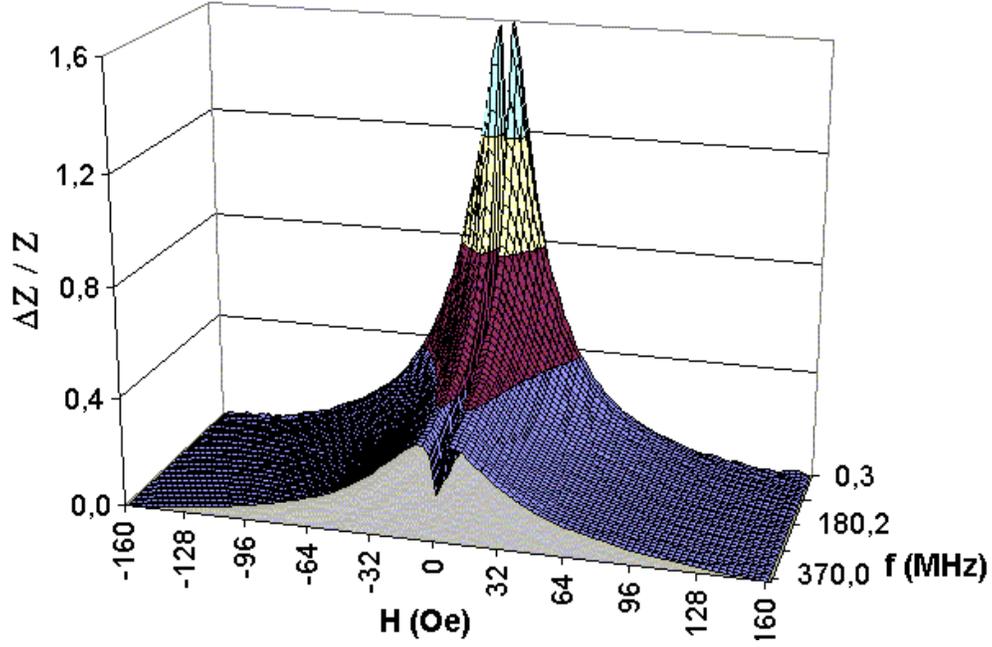} 
\caption{Three-dimensional plot of GMI ratio $\Delta Z/Z$ as a fonction of 
frequency $f$ and magnetic field $H_{dc }$for Vitrovac$^{\textrm{\scriptsize\textregistered}}$
6025 CP-cut sample for the LMI setup.} 
\end{figure}

In the TMI geometry ($H_{dc}$ perpendicular to ribbon axis) with a larger 
$H_{dem }$ field ($ H_{dem } >> H_{kT }$ for high $\mu_{r}$ magnetic material), the 
GMI ratio shows a split-peak Lorentzian-like profile with a separation of 
about 38 Oe between the peaks rounded by the distribution of anisotropy 
field $H_{k}$ (not shown here). Comparison of GMI ratio values for CP and CT 
samples measured at 10 MHz in LMI configuration (Fig. 2) shows a drop of the 
GMI ratio from 160{\%} to 100{\%}. 

\begin{figure}[h!] 
\centering
\includegraphics[angle=0, width=13.5 cm]{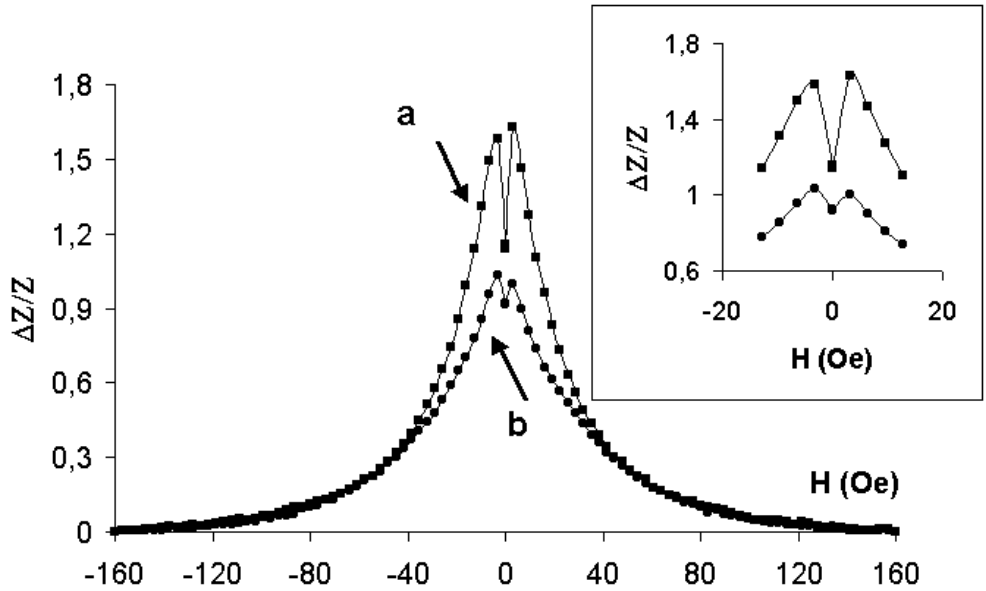} 
\caption{GMI ratio profile $\Delta Z/Z$ as a fonction of magnetic fielf 
$H_{dc}$ in LMI configuration at a frequency of 10MHz for 
Vitrovac$^{\textrm{\scriptsize\textregistered}}$ 6025 CP (a) and CT (b) samples. Inset 
is a zoom-in on the split-peak structure.} 
\end{figure}
 
GMI strongly decreases as the frequency, $f$, increases. The imaginary part of 
impedance \cite{valenzuela} as a function of $H_{dc}$ and $f$ (not shown here) is very similar 
to that of total impedance, indicating that total impedance is made 
essentially of the imaginary part. The real component (not shown here), also 
displays a rounded split-peak shape (with field separation between peaks 
comparable to those of impedance), but is quite insensitive to $f$. This 
behavior can be ascribed to a more direct dependence on the skin effect that 
is expected at the frequencies used.

In conclusion, our work shows the complexity of GMI spectra modulated by the 
dispersion of magnetic anisotropy and influence of the demagnetizing field 
inherent to the geometrical structure of the samples.

\end{document}